\documentclass[a4paper]{jpconf}
\usepackage[cp1252]{inputenc}
\usepackage{iopams}
\usepackage{graphicx}
\usepackage{amssymb}
\usepackage{amsmath}
\usepackage{graphicx}
\usepackage[all]{xy}
\begin{document}
\title{Impact of nuclear gluon distributions on neutrino-induced leptoquark production}

\author{I~Alikhanov$^{1,2}$}

\address{$^1$ Institute for Nuclear Research of the Russian Acad\-emy of Sciences,
Prospekt 60-letiya Oktyabrya 7a, Moscow 117312, Russia}
\address{$^2$ Institute of Applied Mathematics and Automation KBSC RAS, Shortanova 89a, Nalchik 360000, Russia}

\ead{ialspbu@gmail.com}

\begin{abstract}
We analyze neutrino-induced leptoquark production on atomic nuclei. A leptoquark term in the Lagrangian admits the possibility that neutrinos interact with gluons. The current lower limits on the leptoquark masses are of the order of 1~TeV depending on the leptoquark quantum numbers and couplings. Such heavy states can be produced in ultra-high energy cosmic neutrino scattering processes. The four-momentum transfer squared and the Bjorken variable simultaneously probed in these processes may reach values kinematically inaccessible at present collider experiments. We study the impact of the gluon density in a nucleus on the cross section for the leptoquark production.  We show that taking into account the nuclear parton distributions  shifts the production threshold to significantly lower neutrino energies. As a particular case we consider the interaction with oxygen, which is abundant in water/ice neutrino telescopes. 
\end{abstract}

%%%%%%%%%%%%%%%%%%%%%%%%%%%%%%%%%%%%
\section{Introduction\label{introd}}
%%%%%%%%%%%%%%%%%%%%%%%%%%%%%%%%%%%%
Neutrinos coming from outer space to Earth transmit information about their origin, the environment of propagation and the mechanism of oscillations  over cosmological distances and timescales. These particles are perfect messengers for astrophysical and cosmological observations because they are able to travel almost without suffering deflection and absorption~\cite{GalloRosso:2018omb}. On the other hand, the enormous energies that may be possessed by the neutrinos on their way to the detector provide a unique opportunity for testing models of lepton interactions as well as measuring the cross sections in kinematic regions inaccessible at ordinary collider experiments. For example, the IceCube neutrino telescope~\cite{Ahrens:2002dv} has detected neutrino events with energies above 1~PeV ($1~\text{PeV}=10^{15}~\text{eV}$)~\cite{Aartsen:2013bka,Aartsen:2013jdh,Aartsen:2014gkd}. 

A correct interpretation of observations of this kind requires knowledge on the flavor composition of the incoming neutrino flux and on the interaction cross sections. The predicted values of the cross sections at ultra-high energies are subject to the assumptions about the distributions of partons (i.e. quarks
and gluons) in the nucleon~\cite{Gandhi:1998ri,CooperSarkar:2007cv,Connolly:2011vc}. The cosmic ray neutrinos of PeV energies and above probe the parton content of the nucleon at very small values of  the Bjorken variable, $x$, and large four-momentum transfer squared, $Q^2$. The rise of gluon density in this kinematic regime may enhance the neutrino cross section, as was noted in~\cite{McKay:1985nz}. 
In models with leptoquarks neutrinos may interact with gluons. It is therefore interesting to investigate the dependence of the related cross sections on the gluon distributions in the nucleon and atomic nuclei. Leptoquarks are hypothetical particles of spin 0 (scalar) or 1 (vector) that simultaneously carry both the baryon and lepton numbers and may thus mediate lepton and baryon number changing currents. Such states appear in a variety of extensions of the standard electroweak theory as the Pati--Salam model~\cite{Pati:1974yy}, grand unification theories~\cite{Georgi:1974sy,Fritzsch:1974nn}, technicolor schemes~\cite{Lane:1991qh} and in the framework of compositeness of quarks and leptons~\cite{Schrempp:1984nj}. One can restrict the possible quantum numbers of leptoquark states by assuming that their interactions with quarks and leptons are dimensionless and invariant under the $\rm SU(3)_C\times SU(2)_L\times U(1)_Y$ gauge group of the Standard Model~\cite{Buchmuller:1986zs}. 

The prospects for direct observation of leptoquarks at $\rm e^+e^-$,  $\rm ep$, and $\rm pp$ colliders have been studied extensively in the literature~\cite{Buchmuller:1986zs,Djouadi:1989md,Blumlein:1992ej,CiezaMontalvo:1992bs,Aliev:1996qj,Blumlein:1996qp,Kramer:2004df,Ohnemus:1994xf,Alikhanov:2011zf,Alikhanov:2012kk} and corresponding experimental searches have been carried out as well~\cite{Abbiendi:2003iv,Collaboration:2011qaa,Abazov:2011qj,Abramowicz:2012tg,Aaboud:2019bye,Sirunyan:2018kzh,Sirunyan:2018btu}. There are also indirect bounds on leptoquark states obtained from experiments investigating heavy quark decays~\cite{Dorsner:2011ai,Hiller:2018wbv}. The flavor anomalies  found recently in the semileptonic $B$ decays~\cite{Lees:2013uzd,Aaij:2015yra,Hirose:2016wfn} can be explained within leptoquark models~\cite{Becirevic:2018afm}. The current mass limits for leptoquarks from the direct searches are of the order of 1~TeV and depend on their quantum numbers and couplings~\cite{Tanabashi:2018oca}. 

Of special interest is the possibility of leptoquark production induced by ultra-high energy cosmic neutrinos~\cite{Robinett:1987ym,Doncheski:1997it,Anchordoqui:2006wc,Romero:2009vu,Alikhanov:2013fda,Barger:2013pla,Dutta:2015dka,Mileo:2016zeo,Dey:2017ede,Becirevic:2018uab}. In this case, for example, the $\nu_\tau$-component of the astrophysical neutrino flux is concerned~\cite{Barger:2013pla}. It is a challenging problem to isolate tau neutrino events in kilometer-scale detectors~\cite{Cowen:2007ny,Alikhanov:2016qcj} so that this component remains so far unidentified experimentally~\cite{Usner:2017aio}. 

It is a well known
fact that the parton distributions in nuclei differ from those
in the free nucleon due to the many-body
effects in the nuclear medium. This may manifest itself
in the observed cross section
and rapidity distributions~\cite{Goncalves:2001vs,Hirai:2007sx,Adeluyi:2011rt}. In this article evaluate the cross section for the leptoquark production by ultra-high energy neutrinos taking into account the nuclear gluon distributions. To our best knowledge, this has not been done in previous studies. As a particular case, we consider neutrino scattering on oxygen which is abundant in water/ice neutrino telescopes.

The article is organized as follows. In section~\ref{sec_leptoq}, we give the subprocess cross section for single leptoquark production due to neutrino--gluon interaction and briefly explain the details of its derivation. In section~\ref{sec_gluon}, we discuss the gluon distributions in nuclei and propose a parametrization of such a distribution for oxygen. We also describe the technique of calculation of the neutrino--nucleus scattering cross sections. Section~\ref{results_concl} contains numerical results of the calculations. Conclusions are given in section~\ref{concl2}.

%%%%%%%%%%%%%%%%%%%%
\section{The leptoquark model\label{sec_leptoq}}       
\begin{figure}[t]
\centering\includegraphics[width=0.76\columnwidth]{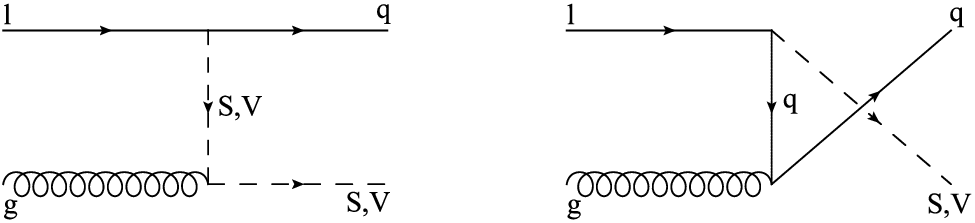}
\caption{Lepton--gluon interaction through leptoquark excitation (${\mathrm l}=\nu_{{\mathrm e}},\nu_{\mu},\nu_{\tau},{\rm e},\mu,\tau$, the symbols ${\mathrm g}$, ${\mathrm q}$, ${\mathrm S}$ and ${\mathrm V}$ denote the gluon, quark, the scalar and vector leptoquarks, respectively.).
\label{fig1}
}
\end{figure}

%%%%%%%%%%%%%%%%%%%%
We adopt the $\rm SU(3)_C \times SU(2)_L \times U(1)_Y$ invariant leptoquark model according to Buchm\"uller, R\"uckl and Wyler~\cite{Buchmuller:1986zs} with the following Lagrangian describing the interaction of left-handed fermions:

\begin{equation}
\mathcal{L}_{\rm I}=\lambda \bar q^c_{\mathrm L}\rmi\tau_2l_{\mathrm L}\mathcal{S}+h\bar q_{\mathrm L}\gamma^\mu l_{\mathrm L}V_\mu+{\mathrm H.c.} \label{main_lag}
\end{equation}
Here $q_{\mathrm L}$ and $l_{\mathrm L}$ are the quark and lepton doublets, respectively, $\tau_2$ is the second Pauli matrix, $\gamma^{\mu=0, 1, 2, 3}$ are the Dirac gamma matrices, $\lambda$ and $h$ are the couplings to the scalar ($\mathcal{S}$) and vector ($V_\mu$) leptoquark fields, respectively, ${\mathrm H.c.}$~means Hermitian conjugation. For example, the interaction in the case of the $\mathrm d$~quark and the tau neutrino will read

\begin{equation}
\mathcal{L}_{\nu\rm I}=-\lambda \bar \psi^c_{\mathrm {d_L}}\psi_{\nu_\tau}\mathcal{S}+h\bar \psi_{\mathrm {d_L}}\gamma^{\mu}\psi_{\nu_\tau}V_\mu+{\mathrm H.c.}, \label{main_lag2}
\end{equation}
where $\psi_{\mathrm {d_L}}$ and $\psi_{\nu_\tau}$ are the Dirac spinors of the left-handed $\mathrm d$ quark and tau neutrino, respectively. One can see that the terms in the Lagrangian have the usual structure, namely the fermion density or current times an external field. In particular, the fermion current $J^{\mu}=\bar\psi_{\mathrm {d_L}}\gamma^{\mu}\psi_{\nu_\tau}$ is contracted with the leptoquark field $V_\mu$, so that the corresponding term, written down briefly, has the familiar form~$hJ^{\mu}V_{\mu}$.  
This model admits the possibility that the  leptons interact with gluons~\cite{Djouadi:1989md}. The lowest order Feynman diagrams contributing to the process

\begin{equation}
{\rm neutrino} + {\rm gluon}\longrightarrow {\rm quark}+{\rm leptoquark} \label{main}
\end{equation}
are shown in figure~\ref{fig1}.

%%%%%%%%%%%%%%
% FIGURE 1
%%%%%%%%%%%%%%
We restrict our consideration to the simplest case of the scalar leptoquark production:

\begin{equation}
 \nu+{\mathrm g}\rightarrow {\mathrm q}+{\mathrm S}.\label{reac_s} 
\end{equation}
Since the neutrino and a fully polarized charged lepton interact with the gluon exactly in the same way, the total cross section for the reaction~\eqref{reac_s} will be given by~\cite{Djouadi:1989md} 

\begin{eqnarray}
\sigma_{\mathrm S}(s)=\frac{\alpha_s\lambda^2}{8s}\left\{\frac{8p\left(m^2_{\mathrm S}-m^2_{\mathrm q}\right)}{s^{3/2}} 
+\ln{\left[\frac{\left(s+2p\sqrt{s}+m^2_{\mathrm q}-m^2_{\mathrm S}\right)^2}{4m^2_{\mathrm S}s}\right]}\right.\nonumber\\\left. 
-4\left(m^2_{\mathrm S}-m^2_{\mathrm q}\right)\frac{s-m^2_{\mathrm S}-m^2_{\mathrm q}}{s^{2}} 
\ln{\left[\frac{\left(s+2p\sqrt{s}\right)^2-\left(m^2_{\mathrm S}-m^2_{\mathrm q}\right)^2}{4m_{\mathrm S}m_{\mathrm q}s}\right]}\right\},\label{sech}
\end{eqnarray}
where $\alpha_s$ is the strong coupling constant,  $\sqrt{s}$ is the invariant energy of the reaction, $m_{\mathrm q}$ and $m_{\mathrm S}$ are the masses of the quark and leptoquark, respectively and $p=\sqrt{\left(s-(m_{\mathrm S}+m_{\mathrm q})^2\right)\left(s-(m_{\mathrm S}-m_{\mathrm q})^2\right)}/(2\sqrt{s})$ is the momentum of a final state particle in the center-of-mass frame. Note that asymptotically, $s\rightarrow\infty$, equation~\eqref{sech} becomes quite simple:

\begin{equation}
\sigma_{\mathrm S}(s)=\frac{\alpha_s\lambda^2}{8s}\ln{\left[\frac{s}{m^2_{\mathrm q}}\right]}.\label{sech_asympt}
\end{equation} 
This is the leading logarithmic limit of the exact cross section.

In the case of the vector leptoquark production there is 
an ambiguity in determination of the coupling to gluons
depending on the nature of the leptoquarks. Considering the reaction $\nu+{\mathrm g}\rightarrow {\mathrm V}+{\mathrm q}$ one will therefore have to make additional assumptions about the form of the leptoquark--gluon vertex~\cite{Alikhanov:2011zf}. The most general $C$ and $P$ conserving coupling is considered in~\cite{Alikhanov:2012kk}.

%%%%%%%%%%%%%%%%%%%%%%%%%%%%%%
\section{Gluon distributions in nuclei\label{sec_gluon}}
%%%%%%%%%%%%%%%%%%%%%%%%%%%%%%
In previous studies of the neutrino-induced leptoquark production only the parton distributions in the nucleon were considered~\cite{Robinett:1987ym,Doncheski:1997it,Anchordoqui:2006wc,Romero:2009vu,Alikhanov:2013fda,Barger:2013pla,Dutta:2015dka,Mileo:2016zeo,Dey:2017ede,Becirevic:2018uab}. For example, the contribution from the gluon content of the nucleon to the leptoquark cross section is comparable to that from quarks, as was recently demonstrated in~\cite{Becirevic:2018uab}.  Meanwhile, targets exploited in neutrino detection experiments frequently consist of nuclei with atomic numbers $A\gtrsim10$. Probably the most common example is water that contains abundant oxygen ($^{16}\text{O}$). It therefore seems to be reasonable to take into account also the parton distributions in the target nucleus as a whole. Since a neutrino--parton scattering subprocess is to occur in this case  typically at the four-momentum transfer squared $Q^2\sim m_{\mathrm S}^2$, the small-$x$ region becomes particularly important and the gluons are also expected to dominate. 
Nuclear gluon distributions have already been used in calculations of the cross sections for various processes in heavy ion  collisions~\cite{Goncalves:2001vs,Hirai:2007sx,Adeluyi:2011rt}.

Thus, we add to the total (per-nucleon) cross section for the leptoquark production a new contribution related to the nucleus: 

\begin{equation}
\sigma_{\mathrm{\nu N}}(E_\nu)=\sigma_{\mathrm N}(E_\nu)+\sigma_{{}^A\mathrm {N}}(E_\nu)/A,\label{sech2p}
\end{equation}
where $E_\nu$ is the incident neutrino energy in the laboratory frame, $\sigma_{\mathrm N}(E_\nu)$ and $\sigma_{{}^A\mathrm {N}}(E_\nu)$ are the cross sections for neutrino--nucleon scattering, $\nu+{\mathrm N}\rightarrow {\mathrm S}+\text{anything}$,  and for neutrino--nucleus scattering, $\nu+{{}^A\mathrm {N}}\rightarrow {\mathrm S}+\text{anything}$.  We focus our attention on the latter process.

Due to ultra-high energies are involved, $\sigma_{i}(E_\nu)$ ($i=\{{\mathrm N},{{}^A\mathrm {N}}\}$)  can be represented, according to the quark--parton model~\cite{Bjorken:1969ja}, as a convolution of the partonic cross section for $\nu+{\mathrm g}\rightarrow {\mathrm S}+{\mathrm q}$ and the relevant gluon distribution:

\begin{equation}
\sigma_i(E_\nu)=\int\limits_{m^2_{\mathrm S}/s}^1\sigma_{\mathrm S}(xs)G_i(x,Q^2)\rmd x,\label{sech2}
\end{equation}
where $G_i(x,Q^2)$ is the nucleonic ($i={\mathrm N}$) or nuclear $(i={}^A\mathrm {N})$ gluon distribution function and $\sigma_{\mathrm S}(s)$ is given by equation~\eqref{sech}, $s=2Am_{\mathrm N}E_\nu$ with the nucleon mass $m_{\mathrm N}=0.938$~GeV. For $i={\mathrm N}$, $s=2m_{\mathrm N}E_\nu$. Note that we  assume $E_\nu\gg m_{\mathrm S}\gg m_{\mathrm q}$.

Following the analysis of~\cite{Hirai:2007sx} we also express the nuclear gluon distribution function in terms of the corresponding nucleonic function:

\begin{equation}
G_A(x,Q^2)=w_{\mathrm g}(x,Q^2,A)f_{\mathrm g}(x,Q^2).\label{distr}
\end{equation}
Here $w_{\mathrm g}(x,Q^2,A)$ indicates the nuclear modification effect, $f_{\mathrm g}(x,Q^2)$ is the gluon distribution function for the free nucleon. The nuclear modification is parametrized as

\begin{equation}
w_{\mathrm g}(x,Q^2,A)=1+\left(1-\frac{1}{A^{1/3}}\right)\frac{a+bx+cx^2+dx^3}{(1-x)^\beta},\label{modif}
\end{equation}
where $a$, $b$, $c$, $d$ and $\beta$ are parameters. In particular, for oxygen $w_{\mathrm g}(x,Q^2,16)\approx 0.9$ at $x=10^{-3}$ and $Q^2=1$~$\text{GeV}^2$. In addition, calculations for $x<10^{-1}$ indicate the tendency $w_{\mathrm g}(x,Q^2,A)\rightarrow1$ as $Q^2$ increases~\cite{Hirai:2007sx}.
Therefore, in this kinematic regime, for simplicity,  we assume  

\begin{equation}
w_{\mathrm g}(x,{Q}^{2},16)=1.\label{modif1}
\end{equation}
Adopting $f_{\mathrm g}(x,Q^2)$, for example, from~\cite{McKay:1985nz} we finally obtain the following functional form for the gluon distribution in the oxygen nucleus: 

\begin{equation}
G_{^{16}\mathrm{O}}(x,Q^2)=K(Q^2)\frac{1}{x}\exp{\left\{\left[\frac{48N_{\mathrm C}}{b}\ln{\left(\frac{1}{x}\right)}\ln{\left(\frac{\ln{\left(Q^2/\Lambda^2\right)}}{\ln{\left(Q_0^2/\Lambda^2\right)}}\right)}\right]^{1/2}\right\}},\label{distr_m}
\end{equation}
where $K(Q^2)$ is a $Q^2$-dependent parameter, $N_{\mathrm C}$ is the number of colors, $b=(11N_{\mathrm C}-2n_{\mathrm f})/3$ with $n_{\mathrm f}$ being the number of flavors, $\Lambda$ is the QCD scale parameter. Carrying over  the properties of the adopted $f_{\mathrm g}(x,Q^2)$ to $G_{^{16}\text{O}}(x,Q^2)$, we assume that equation~\eqref{distr_m} also applies for $10^{-8}<x<10^{-1}$ in the range $10^4~\text{GeV}^2\leq Q^2\leq10^8~\text{GeV}^2$.

%%%%%%%%%%%%%%%%%%%%%%%%%%%%%%%%%%
\section{Numerical results\label{results_concl}}
%%%%%%%%%%%%%%%%%%%%%%%%%%%%%%%%%
%%%%%%%%%%%%%%
% FIGURE 2
%%%%%%%%%%%%%%
\begin{figure}[t]
\centering\includegraphics[width=0.6\columnwidth]{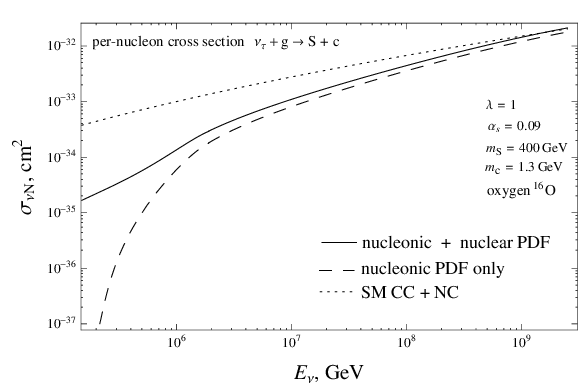}
\caption{The per-nucleon tau neutrino scattering cross sections.  The dashed curve is the contribution from the scalar leptoquark production on gluons in the nucleon obtained in~\cite{Becirevic:2018uab}. The solid curve is the contribution from  the scalar leptoquark production on the gluons in the nucleon plus gluons in the oxygen nucleus as a whole. The dotted curve represents the Standard Model neutrino--nucleon scattering cross section (the sum of the charged current and neutral current contributions). The following values of the parameters are used in the calculations: $\lambda=1$, $\alpha_s=0.09$, $m_{\mathrm S}=400$~GeV, $m_{\mathrm c}=1.3$~GeV. PDF abbreviates a gluon distribution function.
\label{fig2}
}
\end{figure} 
As an example, we have performed a calculation of the per-nucleon cross section for the production of a scalar leptoquark plus a charmed quark, ${\mathrm c}$, in ultra-high energy tau neutrino scattering. In addition to the nucleonic partons, we have also taken into account a possible contribution from the gluons in the oxygen nucleus as a whole.  As far as we know, this is the first evaluation of the contribution  of this gluon content to the neutrino-induced leptoquark production. The obtained numerical results are shown in figure~\ref{fig2}. We have used the parametrization~\eqref{distr_m} in equation~\eqref{sech2} with the following values of the parameters: $\lambda=1$, $\alpha_s=0.09$, $m_{\mathrm S}=400$~GeV, $m_{\mathrm c}=1.3$~GeV, $N_{\mathrm C}=3$, $n_{\mathrm f}=5$ (the role of the ${\mathrm t}$ quark is neglected),  $\Lambda=0.2$~GeV, $Q_0^2=5$~$\text{GeV}^2$, $K(Q^2)=0.025$~\cite{McKay:1985nz}. We have also adopted the static scale $Q^2=m_{\mathrm S}^2$. One can see that there is a good agreement with previous calculations at higher neutrino energies where the nucleonic partons dominate. However, as the neutrino energy decreases, the nuclear effect becomes sizable and significantly shifts the the leptoquark production threshold to lower energies.

 \section{Conclusions\label{concl2}}
Leptoquarks -- scalar or vector bosons -- arise naturally in models unifying the quark and lepton sectors of the Standard Model. Such particles simultaneously carry both the baryon number and the lepton number and thus are able to mediate quark--lepton transitions. They are searched for at electron--positron, electron--proton and hadron--hadron colliders as well as in experiments that study the neutrino component of cosmic rays. There are also indirect bounds on the leptoquark properties coming from the anomalous magnetic moment of the muon and meson decays.  The current lower limits on the leptoquark masses are of the order of 1~TeV depending on the quantum numbers and couplings. 

 A leptoquark term in the Lagrangian admits the possibility that neutrinos interact with gluons. In this article we have studied the impact of the gluon density in a nucleus on the cross section for the neutrino-induced leptoquark production. To our best knowledge, this has been done for the first time.  It is shown that taking into account the nuclear parton distributions  shifts the production threshold to significantly lower neutrino energies. This broadens the energy range where the leptoquark signals may appear. As a particular case we have considered the interaction with oxygen, which is abundant in water/ice neutrino telescopes. 

%%%%%%%%%%%%%%%%%%%%%%
%{\bf Acknowledgements}
%%%%%%%%%%%%%%%%%%%%%%
%%%%%%%%%%%%%%%%%%%%%%%%%%%%%%%%%%%%%%%%%%%%%%%%%%%%%%%%%%%%%%%%%%%%%
\ack
This work was carried out at the Baksan Neutrino Observatory INR RAS (the
Common-Use Center) with financial support of the Ministry of Science and
Higher Education of the Russian Federation: agreement number
075-15-2019-1640, unique identifier of the project RFMEFI62119X0025.
%%%%%%%%%%%%%%%%%%%%%%%%%%%%%%%%%%%%%%%%%%%%%%%%%%%%%%%%%%%%%%%%%%%%%%%%%

\section*{References}
\bibliographystyle{iopart-num}
\bibliography{alikhanov}

\end{document}